\documentclass[pra,twocolumn,floatfix,aps,superscriptaddress,showpacs,amsfonts,10pt]{revtex4}
\usepackage{graphicx}
\usepackage{array}
\usepackage{amsthm}
\usepackage{amsmath}
\usepackage{amssymb}

\begin{document}
\title{Entropic characterization of coherence in quantum evolutions}

\author{Zhengjun Xi}
\email{xizhengjun@snnu.edu.cn}
\affiliation{College of Computer Science, Shaanxi Normal University, Xi'an 710062,
China}

\author{Mingliang Hu}
\affiliation{School of Science, Xi'an University of Posts and Telecommunications, Xi'an 710061, China}

\author{Yongming Li}
\affiliation{College of Computer Science, Shaanxi Normal University, Xi'an 710062,
China}

\author{Heng Fan}
\affiliation{Beijing National Laboratory for Condensed Matter Physics, Institute
of Physics, Chinese Academy of Sciences, Beijing, 100190, China.
}

\affiliation{School of Physical Sciences, University of Chinese Academy of Sciences,
Beijing 100190, China
}

\affiliation{Collaborative Innovation Center of Quantum Matter,
Beijing 100190, China
}

\date{\today}

\begin{abstract}
By using relative entropy of coherence, we characterize the coherence gain induced by some quantum evolutions,
including the cohering power of unitary operations and the decohering power of quantum operations.
We find that the cohering power of the controlled unitary operation can be reduced to the cohering power of
the corresponding unitary operation.
We observe that the global coherence generated via incoherent operation applied to the system
and an incoherent ancilla do not exceed the amount of coherence contained in the initial system.
Our result provides a much tighter lower bound of coherence for the initial quantum state,
and give an interesting chain of inequalities for coherence, quantum correlation and entanglement.
We also strengthen the relations between quantum correlations and coherence.
\end{abstract}
\maketitle
\section{Introduction}
Quantum coherence arising from quantum superposition plays a central role in
quantum mechanics. It is also a common necessary condition for
both entanglement and other types of quantum correlations, and it is also an important physical resource in
quantum computation and quantum information processing.
Quantum coherence represents a basic feature of quantum system.
Recently, quantum coherence quantified in a resource-theoretic framework is proposed~\cite{Aberg06,Baumgratz13,Girolami14},
and it has received a lot of attentions~\cite{Angelo13,Rosario13,Marvian13,Levi14,Marvian14,Karpat14, Aberg14,Monras14,Hai14,Xi14,Xi15,Winter15,Singh15,Lostaglio14,Streltsov15,Chitambar16,Adesso15,Yao15,Yuan15,Pati15,Guo15}.


We know that quantification of quantum coherence depends on a reference basis, and it cannot be increased under
a restricted class of operations known as incoherent operations~\cite{Baumgratz13}.
It is well known that the coherence may change under an unitary operation. In quantum algorithm,
we know the unitary operation could product coherent superposition. Recently, the role of coherence in the Deutsch-Jozsa and related algorithms has been discussed~\cite{Hillery16}.
In quantum information processing, long time quantum coherence is necessary, however,
quantum coherence is fragile and the environmental noises can trigger decoherence.
There is a strong motivation to understand quantitatively how much
coherence may be produced or eliminated in a given operation.
Recently,  the cohering and decohering powers of quantum operations via skew information was studied ~\cite{Mani15}.
Besides skew information, the relative entropy of coherence gives clear physical interpretation and operational interpretation~\cite{Baumgratz13,Xi15,Winter15,Singh15}. Also, the relative entropy of coherence is applied in quantum thermodynamics~\cite{Angelo13,Rosario13,Kammerlande15}.

In this paper, we will consider relative entropy of coherence to discuss the behavior of coherence in quantum evolutions.
As proposed in Ref. ~\cite{Mani15}, the cohering power is defined as the maximum coherence induced by an unitary operation
acting on incoherent states. This definition is similar with the discording power~\cite{Galve13}.
For the decohering power, which is defined as the maximum decay of coherence caused by
quantum channel acting on the maximally coherent state. We find some interesting properties and obtain some the trade-off relations.
As an application, we consider the cohering power of 1-qubit unitary operators and the controlled unitary operator, and the decohering power of the conventional quantum channels on 1-qubit. We also find that the amount of coherence created by some unitary operators acting on coherent states
is more than that of the incoherent states.
This finding motivates us to further discuss the relation between coherence in single system and coherence in the bipartite system.

The paper is organized as follows. In Sec.~\ref{sec:rec}, we introduces the concept of relative entropy of coherence.
In Sec.~\ref{sec:cpuo}, we discuss the cohering power of unitary operations,  In Sec.~\ref{sec:dpqc}, we consider the de-cohering power of quantum channels. In Sec.~\ref{sec:qccoh}, we discuss the relation between coherence in single system and coherence in multipartite system.
We discuss the relations between quantum correlations and coherence in Sec.\ref{sec:qc}.
The conclusions are presented in Sec.~\ref{conclusion}.

\section{Relative entropy of coherence}\label{sec:rec}
In this section, we briefly give an account of the concepts that are required to derive our main results. We will be concerned with the resource theory of coherence as in Ref.~\cite{Baumgratz13}.
Consider a finite $d$-dimensional
Hilbert space $\mathcal{H}$ with a fixed reference basis $\{|i\rangle\}_{i=1}^d$, the set of incoherent states $\mathcal{I}$ is defined as the set of all the states of the form,
$\delta=\sum_i\delta_i|i\rangle\langle i|$,
where $\delta_i$ are probabilities, and $\sum_i\delta_i=1$. Any state which can not be written as above is defined as coherent state,
this implies that coherence is basis-dependent.
Baumgratz \emph{et al.} proposed that any proper measure of the coherence must satisfy some basic conditions~\cite{Baumgratz13}. It has been shown that the relative entropy of coherence and $l_1$-norm of coherence satisfy all those conditions.
Here, we introduce relative entropy of coherence~\cite{Baumgratz13}. For any quantum state $\rho$, the relative entropy of coherence is defined as
\begin{equation}
\mathcal{C}_{\mathrm{re}}(\rho):=\min_{\sigma\in \mathcal{I}}S(\rho||\sigma),
\end{equation}
where $S(\rho||\sigma)=\mathrm{Tr}(\rho\log_2\rho-\rho\log_2\sigma)$ is the relative entropy~\cite{Nielsen}.
As was shown in~\cite{Baumgratz13}, there is an analytical expressions, i.e.,
\begin{equation}\label{eq:rec1}
\mathcal{C}_{\mathrm{re}}(\rho)=S(\Delta(\rho))-S(\rho),
\end{equation}
where $\Delta(\rho):=\sum_i\langle i|\rho|i\rangle|i\rangle\langle i|$ is the diagonal part of the stats $\rho$ in fixed basis, the map $\triangle$ is a completely dephasing channel, and $S(\rho)=-\mathrm{Tr}\rho\log_2\rho$ is the von Neumann entropy~\cite{Nielsen}. Clearly, we have
$\mathcal{C}_{\mathrm{re}}(\rho)\leq S(\Delta(\rho))\leq\log_2d$.
Note that $\mathcal{C}_{\mathrm{re}}(\rho)=S(\Delta(\rho))$ if and only if the quantum state $\rho$ is a pure state.
In particular, if there exists pure states such that $\mathcal{C}_{\mathrm{re}}(\rho)=\log_2d$, these pure states are called maximally coherent states. Baumgratz \emph{et al.} have defined a maximally coherent state, $|\psi\rangle=\frac{1}{\sqrt{d}}\sum_{i=1}^d|i\rangle$.
Note that the maximally coherent state can have the free parameters of phase \cite{PengPRA}.
For example, for the 1-qubit system, let us take the computational basis $\{|0\rangle, |1\rangle\}$, then the states
\begin{equation}
|+\rangle=\frac{1}{\sqrt{2}}(|0\rangle+|1\rangle), |-\rangle=\frac{1}{\sqrt{2}}(|0\rangle-|1\rangle)
\end{equation}
are both maximally coherent states.

\section{Cohering power of unitary operations}\label{sec:cpuo}
\subsection{Definition and Properties}
In Ref.~\cite{Mani15}, the definition of cohering power via skew information is proposed.
In the same way, we can consider cohering power for any coherence measure. Based on the definitions in~\cite{Mani15,Galve13},
for any unitary operation $U$ on Hilbert space $\mathcal{H}$, its cohering power is defined by relative entropy of coherence
\begin{equation}
\mathcal{C}_{\mathrm{re}}(U):=\max_{\delta\in \mathcal{I}}\mathcal{C}_{\mathrm{re}}(U\delta U^\dagger),
\end{equation}
where the maximum is taken over all incoherent states, and $\mathcal{C}_{\mathrm{re}}(\rho)$ is the relative entropy of coherence.
By considering the conditions of the coherence quantification,
we can observe the following inequalities,
\begin{align}\label{eq:upper_bound_1}
\mathcal{C}_{\mathrm{re}}(U\delta U^\dagger)\leq&\ \sum_i\delta_i\ \mathcal{C}_{\mathrm{re}}(U|i\rangle\langle i|U^\dagger)\nonumber\\
\leq&\ \mathcal{C}_{\mathrm{re}}(U|i^\prime\rangle\langle i^\prime|U^\dagger).
\end{align}
where $|i^\prime\rangle$ is one of the fixed reference basis. Thus, we obtain
that the optimization of the cohering power $\mathcal{C}_{\mathrm{re}}(U)$ can be reduced to a simple discrete set via the fixed reference basis $\{|i\rangle\}_{i=1}^d$ on Hilbert space $\mathcal{H}$,
\begin{align}\label{eq:closed exp_cohpower_1}
\mathcal{C}_{\mathrm{re}}(U)=\mathcal{C}_{\mathrm{re}}(U|i^\prime\rangle\langle i^\prime|U^\dagger)
=\max_{i} \mathcal{C}_{\mathrm{re}}(U|i\rangle\langle i|U^\dagger).
\end{align}
This shows that the cohering power of unitary operation is equal to the maximally coherence that is created by performing on the fixed basis.
In other words, the cohering power of a unitary operation can be described by
the amount of the coherence created by unitary operation acting on the computational basis states.
It is natural that we easily give two basic observations as following.

(i) The cohering power is non-negative, and is upper bounded by the dimension of Hilbert space, i.e.,
\begin{equation}\label{Bound_UL}
0\leq \mathcal{C}_{\mathrm{re}}(U)\leq \log_2d.
\end{equation}

(ii) Consider a composite
Hilbert space $\mathcal{H}_A\otimes\mathcal{H}_B$ with a fixed reference product basis $\{|i\rangle^A|j\rangle^B\}_{i=1,j=1}^{d_Ad_B}$, where $\{|i\rangle^A\}_{i=1}^{d_A}$ and $\{|j\rangle^B\}_{j=1}^{d_B}$ are fixed reference basis in $\mathcal{H}_A$ and $\mathcal{H}_B$, respectively,
suppose that $U$ and $V$ are unitary operations on $\mathcal{H}_A$ and $\mathcal{H}_B$, then cohering power is additive,
\begin{equation}
\mathcal{C}_{\mathrm{re}}(U\otimes V)=\mathcal{C}_{\mathrm{re}}(U)+\mathcal{C}_{\mathrm{re}}(V).
\end{equation}
Note that here we take the incoherent basis on the composite system, otherwise, this property in general does not hold.
For example, we take Bell basis as the fixed basis on 2-qubit system, without loss of generality, one takes $\{|0\rangle,|1\rangle\}$ as the fixed basis on 1-qubit system.
In this setting, we consider the cohering power of unitary operation $H\otimes H$, where $H$ is Hadamard operation.

Since $|+\rangle=H|0\rangle$ and $|-\rangle=H|1\rangle$, which are maximally coherent states on the 1-qubit system,
we can easily check that,
\begin{eqnarray}
\mathcal{C}_{\mathrm{re}}(H\otimes H)&=&1,
\nonumber \\
\mathcal{C}_{\mathrm{re}}(H)&=&1.
\end{eqnarray}
This implies that
\begin{equation}
\mathcal{C}_{\mathrm{re}}(H\otimes H)<\mathcal{C}_{\mathrm{re}}(H)+\mathcal{C}_{\mathrm{re}}(H).
\end{equation}

\subsection{The 1-qubit unitary operation}
In this subsection, we will consider unitary operations on 1-qubit system, which is described by a Hilbert space $\mathcal{H}_2$ with the fixed basis $\{|0\rangle,|1\rangle\}$. We know that an arbitrary 1-qubit unitary operation $U$ can be decomposed as a product of three rotations, $R_z(\beta),R_y(\gamma)$ and $R_z(\delta)$, together with a global phase shift$-$a constant multiplier of the form $e^{i\alpha}$~\cite{Nielsen},
\begin{equation}
U=e^{i\alpha}R_z(\beta)R_y(\gamma)R_z(\delta),
\end{equation}

where $\alpha$, $\beta$, $\gamma$ and $\delta$ are real-valued parameters. This is to say, the 1-qubit unitary operation $U$ depends on these four parameters, then we can denote $U(\alpha,\beta,\delta,\gamma)$ for arbitrary 1-qubit unitary operation. Note that the rotations $R_z(\beta)$ and $R_z(\delta)$ do not produce the coherence. Then, the cohering power of the unitary operation $U(\alpha,\beta,\gamma,\delta)$ depends only on the rotation $R_y(\gamma)$. Using the closed expression~(\ref{eq:closed exp_cohpower_1}), we easily check that
\begin{align}
\mathcal{C}_{\mathrm{re}}(U(\alpha,\beta,\gamma,\delta))=H\left(\cos^2\frac{\gamma}{2},\sin^2\frac{\gamma}{2}\right),
\end{align}
where $H(x,y)=-x\log_2x-y\log_2y$ is the binary Shannon entropy.
Thus, we find that the upper bound and lower bound~(\ref{Bound_UL}) can be saturated for Hadamard operation and Pauli operations, respectively.

\subsection{The controlled unitary operation}
It is known that the controlled operation is one of the most useful gates in computing, for both cases of classical and quantum. In quantum computing, the prototypical controlled operation is the CNOT in 2-qubit system. In the following, we discuss the arbitrary controlled unitary operation. Suppose we have $n+k$ qubits, and $U$ is a unitary operator acting on $k$ qubits. Then, we take a fixed product basis $\{|x_1\cdots x_nx_{n+1}\cdots x_{n+k}\rangle\}$ for $n+k$-qubit system, where $x_i\in\{0,1\}$,
the controlled unitary operation $C^n_U$ is a $n+k$-qubit unitary operation, again with $n$ control qubits and $k$ target qubits. If the control qubits are all equal to one,  then $U$ is applied to the target qubits, otherwise the target qubits are left alone. That is,
\begin{equation}
C^n_U|x_1\cdots x_n\rangle|\psi\rangle=|x_1\cdots x_n\rangle U^{x_1\cdots x_n}|\psi\rangle,
\end{equation}
where $x_1\cdots x_n$ in the exponent of $U$ means the product of the bits $x_1, \cdots, x_n$.
Since the relative entropy of coherence is additive, we then have
\begin{equation}
\mathcal{C}_{\mathrm{re}}(C^n_U|x_1\cdots x_n\rangle|\psi\rangle)=\mathcal{C}_{\mathrm{re}}(U^{x_1\cdots x_n}|\psi\rangle).\label{C_U_n}
\end{equation}
Thus, for any controlled unitary operation, its cohering power can be reduced to the coherent power of $k$-qubits unitary operator $U$.
That means,
\begin{align}
\mathcal{C}_{\mathrm{re}}(C^n_U)=&\ \max_{x_{n+1}\cdots x_{n+k}}\{\mathcal{C}_{\mathrm{re}}(U^{x_1\cdots x_n}|x_{n+1}\cdots x_{n+k}\rangle)\}\nonumber\\
=&\ \mathcal{C}_{\mathrm{re}}(U).
\end{align}
This result shows that the cohering power of the controlled unitary operation can be determined by the unitary operator on the target qubits.
In particular, for the $\mathrm{CNOT}$ operation, it is not possible to create coherence from fixed incoherent basis, it only changes one computational state to another one, that is,
$\mathrm{CNOT}|i\rangle|j\rangle=|i\rangle|i\oplus j)\rangle$.
As we already know that $\mathrm{CNOT}$ gate is an incoherent operation for two-qubit system.
But we know that $\mathrm{CNOT}$ operation can be used to create entanglement~\cite{Streltsov15}, e.g.,
\begin{equation}\label{2-qubit_qcc}
\mathrm{CNOT}|+\rangle|0\rangle=\frac{1}{\sqrt{2}}|00\rangle+|11\rangle.
\end{equation}
It is easy to check that
$\mathcal{C}_{\mathrm{re}}(\mathrm{CNOT}|+\rangle|0\rangle)=\mathcal{C}_{\mathrm{re}}(|+\rangle)=1$.
This shows that the local coherence is perfectly changed into the global entanglement by an incoherent operation, meanwhile, the local states become incoherent states. In other words, if once the system possesses quantum coherence, it is possible to entangle it with the outside world under the noise environment.

\section{Decohering power of quantum channel}\label{sec:dpqc}
In this section, we will consider the decohering power of quantum channel. We denote that
$\mathcal{M}$ is a set of the maximally coherent states. Following the definition~\cite{Mani15}, we can define the decohering power of a quantum channel $\cal{E}$ as the maximal coherent difference between before and after quantum channel acting on the maximally coherent states.
Suppose $\cal{E}$ is a quantum channel, then the decohering power of quantum channel $\cal{E}$ is defined as
\begin{equation}\label{def:decohering power}
{\mathcal{D}_{\mathrm{re}}}(\cal{E}):
=\max_{\rho\in\mathcal{M}}\left[{\mathcal{C}_{\mathrm{re}}}(\rho)-{\mathcal{C}_{\mathrm{re}}}(\cal{E}(\rho))\right],
\end{equation}
where the maximum is taken over all maximally coherent states.
Since all maximally coherent states are pure states, then the above definition can be written as
\begin{equation}\label{def:decohering power_2}
{\mathcal{D}_{\mathrm{re}}(\cal{E})}
=\log_2d-\min_{|\psi\rangle\in\mathcal{M}}\mathcal{C}_{\mathrm{re}}(\cal{E}(|\psi\rangle)).
\end{equation}
From the result in~\cite{Xi15}, for any state $|\psi\rangle$, we have
\begin{equation}\label{eq:ur_1}
\mathcal{C}_{\mathrm{re}}({\cal{E}}(|\psi\rangle))+S({\cal{E}}(|\psi\rangle)) \leq \log_2d.
\end{equation}
Substituting the inequality~(\ref{eq:ur_1}) into~(\ref{def:decohering power_2}), we then obtain
\begin{equation}
{\mathcal{D}_{\mathrm{re}}(\cal{E})}\geq \max_{|\psi\rangle\in\mathcal{M}}S(\cal{E}(|\psi\rangle)).
\end{equation}
If the quantum channel satisfies $[{\cal{E}}, \Delta]=0$ and ${\cal{E}}(I/d)=I/d$, we have
\begin{equation}
{\mathcal{D}_{\mathrm{re}}(\cal{E})}=\max_{|\psi\rangle\in\mathcal{M}}S(\cal{E}(|\psi\rangle)).
\end{equation}
 From the model of quantum channel, we know that the entropy $S(\cal{E}(|\psi\rangle))$ could describe the amount of the entanglement between the decayed  system and the outside world. This is to say, the decohering power of quantum channel
is no less than the entangling capability. In other words, the more the entangling capability, the larger the decohering power.

Next, we consider the composite quantum system $AB$. Suppose that $\cal{E}$ and $\Phi$ are quantum channels on quantum systems $A$ and $B$, respectively, let $\mathcal{M}^{AB}$ be the set of the maximally coherent states on $AB$,
then, we have,
\begin{align}
{\mathcal{D}_{\mathrm{re}}({\cal{E}}\otimes \Phi)}=&\ \log_2{d_Ad_B}-\min\mathcal{C}_{\mathrm{re}}(({\cal{E}}\otimes \Phi)(|\psi^{AB}\rangle))\nonumber\\
\geq &\ \log_2{d_Ad_B}-\min\mathcal{C}_{\mathrm{re}}(({\cal{E}}\otimes \Phi)(|\psi^{A}\rangle\otimes|\psi^{B}))\nonumber\\
=&\ \mathcal{D}_{\mathrm{re}}({\cal{E}})+{\mathcal{D}}_{\mathrm{re}}(\Phi),
\end{align}
where the inequality comes from the fact that the product of local maximally coherent states could not construct all global maximally coherent states,
and $|\psi^{A}\rangle$ and $|\psi^{B}\rangle$ are the maximally coherent states on $A$ and $B$, respectively.

We consider quantum channels on 1-qubit system. We know that the maximally coherent states are $|+\rangle$ and $|-\rangle$.
For quantum channels of bit flip, phase flip, bit-phase flip, amplitude damping and depolarizing, we easily find that the decoherence of the maximally coherent states are the same. Then, for these quantum channels, we have,
\begin{equation}
{\mathcal{D}_{\mathrm{re}}(\cal{E})}+\mathcal{C}_{\mathrm{re}}({\cal{E}}(|+\rangle))=1.
\end{equation}
Note that the decohering power of the bit flip channel is zero, this is because that the state $|+\rangle$ on the $x$ axis is left alone via the effect of the bit flip channel on the Bloch sphere.

\section{local coherence convert global coherence via incoherent operations}\label{sec:qccoh}
In this section, we will consider the relationship between local coherence and global coherence. But before proceed, we introduce the dynamics of an open system $A$, which is regarded as arising from an interaction between the system of interest and an environment $E$,
where they form a closed quantum system $AE$. Suppose that the system-environment input state is a product state, $\rho^A\otimes |0\rangle^E\langle 0|$, the quantum operation $\cal{E}$ describing this process is
\begin{equation}\label{eq:q_oper_model}
\mathcal{E}(\rho^A)=\mathrm{Tr}_E(U\rho^A\otimes |0\rangle^E\langle 0|U^\dagger),
\end{equation}
where $U$ is an unitary operation on $AE$.

Consider an incoherent states (or the computational basis) $\{|00\rangle,|01\rangle,|10\rangle,|11\rangle\}$ of the system $AE$,  we know that the CNOT operation is one of the unitary interaction of
phase damping channel~\cite{Nielsen}, and it is not possible to create coherence
from this incoherent states, but it is used to create entanglement from coherent state, seen Eq.~(\ref{2-qubit_qcc}). Under this basis, the unitary operator $U_{adc}$ of amplitude damping channel is~\cite{Giovannetti05}
\begin{equation}
U_{adc}=\left(
  \begin{array}{cccc}
    1 & 0 & 0 & 0 \\
    0 & \sqrt{\eta} &  \sqrt{1-\eta} & 0 \\
    0 & - \sqrt{1-\eta} &  \sqrt{\eta} & 0 \\
    0 & 0 & 0 &1 \\
  \end{array}
\right),
\end{equation}
where $\eta\in[0,1]$. We can easily check that
\begin{align}
U_{adc}|00\rangle=&|00\rangle,\nonumber\\
U_{adc}|11\rangle=&|11\rangle,\nonumber\\
U_{adc}|01\rangle=&\sqrt{\eta}|01\rangle-\sqrt{1-\eta}|10\rangle,\nonumber\\
U_{adc}|10\rangle=&\sqrt{1-\eta}|01\rangle+\sqrt{\eta}|10\rangle.
\end{align}
This shows that the unitary operator $U_{adc}$ is not only possible to create coherence, but also it is possible to create entanglement from some bipartite incoherent states. Further, this operation is not incoherent operation in 2-qubits system. But it can induce an incoherent operation in 1-qubit system. We also find the amount of coherence created by coherent state is more than the incoherent state. To check the case, we consider that the unitary operator $U_{adc}$ act on the local coherent state $|+\rangle|0\rangle$,
\begin{equation}
U_{adc}|+\rangle|0\rangle=\frac{1}{\sqrt{2}}|00\rangle+\frac{1}{\sqrt{2}}U_{adc}|10\rangle.
\end{equation}
After simple calculation, we have that
\begin{align}
\mathcal{C}_{\mathrm{re}}(U_{adc}|+\rangle|0\rangle)=1+\frac{1}{2}\mathcal{C}_{\mathrm{re}}(U_{adc}).
\end{align}
Based on this fact, we consider a sup-cohering power of a bipartite unitary coherent operation induced by quantum channel $\cal{E}$, which is defined by the maximum coherence caused by acting on local coherent states, that is,
\begin{equation}
\mathcal{C}_{\mathrm{re}}(U_{\cal{E}}):=\max\mathcal{C}_{\mathrm{re}}(U_{\cal{E}}|\psi\rangle|0\rangle),
\end{equation}
where the maximization is taken over all the states, which have the form $|\psi\rangle|0\rangle$ with $|\psi\rangle$ is maximal coherent state on local system, and $|0\rangle$ is incoherent state on the other local system. For example, for 2-qubits system, the set $\{|+\rangle|0\rangle, |-\rangle|0\rangle,|+\rangle|1\rangle,|-\rangle|1\rangle\}$ is a local coherent basis.
Thus, using the properties of relative entropy of coherence, we obtain an uncertainly relation as following, that is,
\begin{equation}
\mathcal{D}_{\mathrm{re}}({\cal{E}})+\mathcal{C}_{\mathrm{re}}(U_{\cal{E}})\geq \log_2d,
\end{equation}
where $U_{\cal{E}}$ is an unitary interaction operation of quantum channel $\cal{E}$.

It is shown that any nonzero amount of coherence in a system $A$ can be converted to entanglement between $A$ and an initially incoherent ancilla $E$, by means of incoherent operations~\cite{Streltsov15}. This finding leads to a novel method to quantify coherence in terms of entanglement.
Afterwards, some other research also studied the amount of coherence which can be converted to quantum correlation via incoherent operation~\cite{Ma15,Killoran16}. Those results give some bounds on the relationship between coherence and quantum correlations.
We will try to strengthen some of the results.

Suppose that the state $\rho^A$ of the system $A$ interacts with an ancilla $E$ initialized in a reference incoherent state $|0\rangle^E\langle 0|$ via incoherent operation $\Lambda^{AE}$, after this, the final state $\Lambda^{AE}[\rho^S\otimes|0\rangle^A\langle 0|]$. Based on the properties of relative entropy of coherence, we have
\begin{align}\label{low_bound_coh}
\mathcal{C}_{\mathrm{re}}(\rho^A)
=&\ S\left(\rho^A\otimes|0\rangle^E\langle 0|\parallel|\Delta(\rho^A)\otimes|0\rangle^E\langle 0|\right)\nonumber\\
\geq&\ S\left(\Lambda^{AE}[\rho^A\otimes|0\rangle^E\langle 0|]\parallel\Lambda^{AE}[\Delta(\rho^A)\otimes|0\rangle^E\langle 0|]\right)\nonumber\\
\geq&\ \mathcal{C}_{\mathrm{re}}\left(\Lambda^{AE}[\rho^A\otimes|0\rangle^E\langle 0|]\right).
\end{align}
The first inequality come from the fact the relative entropy is non-increasing under quantum operations, and the second inequality is because the incoherent operation $\Lambda^{AE}$ can not product coherence on the incoherent state $\rho_{\mathrm{diag}}^A\otimes|0\rangle^E\langle 0|$.

For simplicity, we denote $\rho^{AE}:=\Lambda^{AE}[\rho^A\otimes|0\rangle^E\langle 0|]$. We say that the measure $\mathcal{C}_{\mathrm{re}}(\rho^{AE})$ describe the amount of coherence in the composite system $AE$, we call it the global coherence, which is defined by
\begin{equation}
\mathcal{C}_{\mathrm{re}}(\rho^{AE}):=\min_{\zeta^{AE}\in GI} S(\rho^{AE}||\zeta^{AE}),
\end{equation}
where the minimization is taken over the set of $GI$ bipartite incoherent states. Namely, all such states have the following form~\cite{Streltsov15},
$\zeta^{AE}=\sum_ip_i\sigma^A_i\otimes\sigma^E_i$,
where $p_i$ are probabilities and the states $\sigma^A_i$ and $\sigma^E_i$ are incoherent states in the subsystem $A$ and $E$, respectively.

From the results~\cite{Streltsov15,Ma15,Killoran16,Chitambar16},  the inequality~(\ref{low_bound_coh}) provides a new lower bound of initial coherence.
In other words, the amount of global coherence generated via an incoherent operation cannot exceed the amount of initial coherence.
In addition, we can obtain an interesting chain for coherence, entanglement and quantum correlations, i.e.,
\begin{align}\label{eq:chain_relation_1}
\mathcal{C}_{\mathrm{re}}(\rho^A)\geq \mathcal{C}_{\mathrm{re}}(\rho^{AE})\geq Q_{\mathrm{re}}(\rho^{AE})\geq E_{\mathrm{re}}(\rho^{AE}).
\end{align}
Here, we denote $\rho^{AE}:=\Lambda^{AE}[\rho^A\otimes|0\rangle^E\langle 0|]$, and quantum correlation $Q_{\mathrm{re}}$ and $E_{\mathrm{re}}$ are defined by~\cite{Modi10,Vderal97}
\begin{equation}
Q_{\mathrm{re}}=\min_{\xi\in C} S(\rho||\xi),\ \ \  E_{\mathrm{re}}=\min_{\sigma\in S} S(\rho||\sigma),
\end{equation}
where $C$ is the set of classical states and $S$ is the set of separable states.
In addition, for the generalized CNOT gate~\cite{Streltsov15}, we also obtain the following relations,
\begin{equation}\label{eq:qc_coh_2}
\mathcal{C}_{\mathrm{re}}(\rho^A)= \mathcal{C}_{\mathrm{re}}(\rho^{AE})=Q_{\mathrm{re}}(\rho^{AE})=E_{\mathrm{re}}(\rho^{AE}).
\end{equation}

\section{The relations between quantum correlations and coherence}\label{sec:qc}
From the previous section, we know that the creation of coherence with bipartite incoherent operations is bounded by the amount of coherence consumed in its subsystems during the process. In this section, we will clarify the relation quantum correlations and coherence.
For the bipartite quantum system $AB$ with fixed incoherent basis $\{|i\rangle^A|j\rangle^B\}_{i=1,j=1}^{d_Ad_B}$, we denote bipartite quantum states as follows
$$\rho^{AB}=\sum_{i,i^\prime,j,j^\prime}\rho_{i,i^\prime,j,j^\prime}|i\rangle^A\langle i|\otimes |i\rangle^B\langle i|.$$
Without loss of generality, Alice and Bob process $\rho^{A}=\sum_{i,i^\prime}\rho_{i,i^\prime}|i\rangle^A\langle i|$ and $\rho^{B}=\sum_{j,j^\prime}\rho_{j,j^\prime}|j\rangle^B\langle j|$, respectively. To extract information, Alice can perform projective measurement $\{|i\rangle\langle i|\}$ on her party, then the quantum state $\rho^{AB}$ becomes a classical-quantum state or quantum-incoherent state~\cite{Chitambar16}
\begin{equation}
\chi^{AB}=\sum_{i}p_i|i\rangle^A\langle i|\otimes\rho^B_i,
\end{equation}
where $\rho^B_i=\sum_{j,j^\prime}\rho_{i,i,j,j^\prime}|j\rangle^B\langle j^\prime|/p_i$ is the remaining state of $B$ after obtaining the outcome $i$ on $A$ with the probability $p_i=\mathrm{Tr}(\sum_{j,j^\prime}\rho_{i,i,j,j^\prime}|j\rangle^B\langle j^\prime|)$.
For the local measurement can be used to remove the coherence in the measured subsystem, meanwhile it destroys quantum correlations contained in the bipartite quantum system. In that case, we obtain that the sum between
the amount of quantum correlation destroyed by local measurements and the amount of coherence contained in the measured system is equal to the QI relative entropy,
that is,
\begin{equation}\label{D_C_1}
\mathcal{C}_{\mathrm{re}}(\rho^A)+\delta^\rightarrow(\rho^{AB}|\{|i\rangle\langle i|\})=\mathcal{C}^{\rightarrow}_\mathrm{re}(\rho^{AB}).
\end{equation}
Here $\delta^\rightarrow(\rho^{AB}|\{|i\rangle\langle i|\})$ is the discord-like quantity,
and here we do not consider the optimization for the definition of discord.  The QI relative entropy $\mathcal{C}^\rightarrow_\mathrm{re}(\rho^{AB})$ is defined as
$$\mathcal{C}^\rightarrow_\mathrm{re}(\rho^{AB})=\min_{\chi^{AB}\in QI}S(\rho^{AB}||\chi^{AB}),$$
where the minimization is taken over all the quantum incoherent states~\cite{Chitambar16}. The relation~(\ref{D_C_1}) was given in Ref.\cite{Xi15}, we will strengthen the relation by considering the optimization as follows.
Firstly, we can obtain interesting relations between quantum correlations and coherence, i.e.,
\begin{equation}
\mathcal{C}_{\mathrm{re}}(\rho^A)+\delta^\rightarrow(\rho^{AB})\leq\mathcal{C}^\rightarrow_\mathrm{re}(\rho^{AB}).
\end{equation}
Here, the discord $\delta^\rightarrow(\rho^{AB})$ describes quantum correlations in the bipartite quantum system~\cite{Zurek01}.
In fact, the one-way quantum deficit is the minimum amount of the QI relative entropy over all basis of $A$, that is,
\begin{equation}
\Delta^\rightarrow(\rho^{AB})=\min \mathcal{C}^\rightarrow_\mathrm{re}(\rho^{AB}),
\end{equation}
where the minimization is taken over all basis of $A$. After some algebraic manipulation, compared with the results~\cite{Xi15}
where the optimal measurements should be different,
we obtain a compact and reasonable relation, i.e.,
\begin{equation}
\mathcal{C}_{\mathrm{re}}(\chi^{AB})+\mathcal{C}^\rightarrow_{\mathrm{re}}(\rho^{AB})=\mathcal{C}_{\mathrm{re}}(\rho^{AB}).
\end{equation}

Secondly, let the fixes basis $\{|i\rangle\langle i|\}$ on $A$ be the optimal projective measurement for the one-way quantum deficit, then we have
\begin{equation}
\mathcal{C}_{\mathrm{re}}(\rho^A)+\delta^\rightarrow(\rho^{AB})\leq\Delta^\rightarrow(\rho^{AB}).
\end{equation}
Similarly, let the fixes basis $\{|i\rangle\langle i|\}$ on $A$ be the optimal projective measurement for the discord, then we have
\begin{equation}
\Delta^\rightarrow(\rho^{AB})\leq \mathcal{C}_{\mathrm{re}}(\rho^A)+\delta^\rightarrow(\rho^{AB}).
\end{equation}
Therefore, if the fixes basis $\{|i\rangle\langle i|\}$ on $A$ is also the optimal projective measurement for the two, we then obtain a strong trade-off as follows,
\begin{equation}
\Delta^\rightarrow(\rho^{AB})=\mathcal{C}_{\mathrm{re}}(\rho^A)+\delta^\rightarrow(\rho^{AB}).
\end{equation}




\section{Conclusion}\label{conclusion}
We have discussed the cohering power of the unitary operation and the decohering power of quantum channel. For the former, we found that it depends only on the fixed reference basis, and it also satisfies additivity relation. For the latter,
it is sub-additive, and the decohering power of quantum channel
is no less than the entangling capability caused by the channel interacted with the outside world.
We have obtained that the decohering power of 1-qubit quantum channel depends only on the maximally coherent state $|+\rangle$. We have proved that the global coherence generated via incoherent operation applied to the system and an incoherent ancilla could do not exceed the amount of coherence contained in the initial system. Finally, we have improved and strengthen the results in~\cite{Xi15}, and obtained some new relations between quantum correlations and coherence. We hope that
the results presented in this paper are also useful for this
more general scenario.

\section{Acknowledgments}
Z. Xi is supported by NSFC (Grant Nos. 61303009, 11531009), and Fundamental Research Funds for the Central Universities (GK201502004). M.Hu is supported by NSFC (Grant No. 11205121). Y. Li is supported by NSFC(Grant No.11271237). H. Fan is supported by MOST of China (2016YFA0302104), NSFC (Grant No. 91536108),
CAS (XDB01010000, XDB21030300). 


\end{document}